\documentclass[12pt]{article}

\usepackage[utf8]{inputenc}
\usepackage[T1]{fontenc}
\usepackage{lmodern}

\usepackage{amsmath, amssymb, bm}

\usepackage{graphicx}
\usepackage[colorlinks=true, linkcolor=blue, citecolor=blue, urlcolor=blue]{hyperref}

\usepackage[margin=1in]{geometry}

\title{Mode-dependent Emission Thresholds and Frequencies in Metal Nanoparticles within Gain-enhanced Media}

\author{
Karen Caicedo$^{1,2}$,
Nicole Recalde$^1$,
Milena Mora$^1$,\\
Alonso Moreta$^1$,
Maria Antonia Iatì$^3$,
Onofrio M. Maragò$^3$,\\
Melissa Infusino$^{1,3}$,
Alessandro Veltri$^{1,3}$\thanks{Corresponding author: \texttt{aveltri@usfq.edu.ec}}
}

\date{}

\begin{document}
\maketitle

\begin{center}
{\small
$^1$ Colegio de Ciencias e Ingeniería, Universidad San Francisco de Quito, Quito, Ecuador \\
$^2$ Institute of Applied Sciences and Intelligent Systems, National Research Council, Napoli, Italy \\
$^3$ Institute for Chemical and Physical Processes, National Research Council, Messina, Italy
}
\end{center}

\vspace{1em}
 
\begin{abstract}
We present a mode-resolved analysis of emission thresholds in metal nanoparticles embedded within an infinite gain medium, using Mie scattering theory as a rigorous framework. Focusing on the first three resonant modes, we identify the onset of emission by locating the complex zeros of the electric scattering coefficient $a_n(\omega)$, which serve as indicators of lasing conditions. Our hybrid numerical scheme—combining coarse bracketing and two-dimensional Newton refinement—enables precise determination of both the threshold gain $G^\mathrm{th}_{\bullet\, n}$ and corresponding emission frequency $\omega^\mathrm{th}_{\bullet\, n}$ across varying particle radii. The results uncover strong size-dependent behavior, with higher-order modes in larger particles exhibiting significantly reduced threshold gain. These insights elucidate the modal dynamics underlying plasmonic nanolasing and provide design guidelines for active photonic systems leveraging gain-enhanced nanoparticles.
\end{abstract}

\vspace{1em}

\section{Introduction}

The study of plasmonic resonances in metal nanoparticles, particularly when coupled with gain media, has been a vibrant field of research for several decades. A major focus has been the compensation of inherent losses in plasmonic systems~\cite{Berini:2012,Hess:2012,DeLuca:2012,Meng:2011,Infusino:2014,Qian:2017,Polimeno:2020,ShiliangQu:2015-2} and the development of emissive nano-devices, such as Surface Plasmon Amplification by Stimulated Emission of Radiation (SPASER) and nanoscale lasers~\cite{ber03,Noginov:2009,Oulton:2009,Ellis:2024,Ma:2021,Azzam:2020,Stockman:2020,Wu:2019,Xu:2019,Ma:2019,Yang:2017,Deeb:2017,Wang:2017}. In this context, Mie Scattering Theory has provided a robust theoretical framework for modeling the optical response of spherical metal nanoparticles~\cite{rahaman:2016,monreal:2014,gopinath:2008,mishchenko:2019,tcherniak:2010,pathak:2016,fujii:2014}.

More recently, a new paradigm in nanophotonics—known as {\em Mie-tronics}—has emerged. This rapidly evolving framework extends the traditional use of Mie Theory toward the design and engineering of nanoscale devices based on the modal structure of Mie resonances, with applications ranging from sensing to light generation and manipulation~\cite{Kivshar:2022,Won:2019,Rybin2024Metaphotonics}.

Despite the increasing interest in active plasmonic systems, the systematic application of Mie Theory to metal nanoparticles embedded in gain media remains limited. While a few studies have explored this direction~\cite{kulishov:2006,Recalde:2023,Zhu:2015,Shen:2016}, most of the existing characterizations of emission thresholds and gain coupling are confined to the Quasi-Static (QS) approximation~\cite{Veltri:2009,Caicedo:2022}. Although analytically convenient, the QS approximation is strictly valid only for small nanoparticles and primarily describes dipolar responses, thus leaving the behavior of higher-order modes largely unexplored.

In this study, we go beyond the QS regime and extend the application of Mie Scattering Theory to analyze the mode-specific emission characteristics of metal nanoparticles in an infinite gain medium. Specifically, we explore the parameter space for the first three modes in a model system constituted of a silver spherical nanoparticle hosted in gain-enhanced water. By systematically examining the threshold gain and emission frequencies across different particle sizes and modes, we aim to provide a more complete characterization of these phenomena. Notably, our findings indicate that the threshold gain required to drive emission tends to be lower for higher-order modes. This opens new possibilities for the design of emissive plasmonic devices that capitalize on these higher modes, offering a novel approach to achieving efficient plasmonic emission.

\subsection*{Theoretical Background}

In the quasi-static (QS) regime, it has been shown that when a parameter $G$, representing the strength of optically pumped gain in the medium, exceeds a threshold value $G^\mathrm{th}$, the system undergoes a transition to emission. This transition corresponds to the appearance of a singularity in the complex-valued polarizability of the nanoparticle at a specific frequency $\omega^\mathrm{th}$~\cite{Veltri:2012}. The threshold value $G^\mathrm{th}$ depends on the central frequency at which gain is introduced into the system, denoted $\omega_g$ and referred to as the \emph{gain center frequency}. This emission criterion, which depends sensitively on the choice of $\omega_g$, has been validated dynamically through a multipolar eigenmode analysis~\cite{Veltri:2016}, and has also been successfully extended to more complex nanostructures, including core-shell and nano-shell geometries~\cite{Caicedo:2022}
\footnote{While the terms are sometimes used interchangeably, we refer to “core–shell” structures as metal–dielectric bilayers, and “nanoshells” specifically to metallic shells surrounding a dielectric core.}.

In these earlier works, it was found that the threshold gain $G^\mathrm{th}$ required to trigger emission is minimized when the gain spectrum is centered at the zero-gain plasmon resonance frequency, denoted $\omega^0$ (i.e., when $\omega_g = \omega^0$). Under these optimal conditions, the emission frequency coincides with the zero-gain resonance, such that $\omega^\mathrm{th} = \omega_g = \omega^0$. If the gain center is detuned from the resonance, emission can still occur, but only at the cost of a higher threshold gain, and the emission frequency $\omega^\mathrm{th}$ appears between the gain center and the zero-gain resonance~\cite{Veltri:2012}.

Beyond the QS approximation, the role of the polarizability is taken up by the electric Mie scattering coefficients $ a_n $, each associated with a specific multipolar mode. In a recent numerical study~\cite{Recalde:2023}, it was demonstrated that for a spherical particle embedded in an infinite gain medium, the onset of emission is signaled by a singularity in the complex-valued coefficient $ a_n(\omega) $. The gain threshold $ G^\mathrm{th}_n $ is defined as the minimum amount of gain required to induce a singularity in the Mie coefficient $ a_n $, under the condition that the gain center frequency is appropriately tuned to favor emission for the mode $ n $.

Due to the algebraic complexity of the Mie scattering coefficients~\cite{bohren98}, determining the threshold gain and emission frequency analytically is highly nontrivial. In this work, we develop and implement a dedicated numerical protocol to identify the critical values of gain $G^\mathrm{th}_n$ and frequency $\omega^\mathrm{th}_n$ at which each coefficient $a_n$ becomes singular, thereby signaling the onset of emission for the corresponding mode. This iterative algorithm, based on a refined bisection method (see Methods), represents a central contribution of our study. In particular, it allows for a systematic and flexible exploration of how the emission characteristics depend on the particle radius—through the effective size parameter $ x = 2\pi n_m r / \lambda $ (here $ n_m $ is the refractive index of the surrounding medium and $ \lambda $ is the incident wavelength)—and on the specific mode $ n $ targeted for amplification via gain.

Interestingly, while—as in the QS regime—the critical values $G^\mathrm{th}_n$ and $\omega^\mathrm{th}_n$ still depend strongly on the chosen gain center frequency $\omega_g$, and $G^\mathrm{th}_n$ still tends to increase for values of $\omega_g$ far from $\omega^0_n$ (the frequency that maximizes $|a_n|^2$ in the absence of gain), we find that $\omega^0_n$ does not coincide with either the optimal gain center frequency or the corresponding emission frequency. Instead, the following set of quantities naturally arises:
\begin{itemize}
    \item $\omega^0_n$: the zero-gain resonance frequency, i.e., the frequency at which $|a_n|^2$ is maximal for $G = 0$;
    \item $G^\mathrm{th}_{\bullet\, n}$: the minimum threshold gain required to induce emission for mode $n$;
    \item $\omega^g_{\bullet\, n}$: the optimal gain center frequency, i.e., the value of $\omega_g$ that minimizes $G^\mathrm{th}_n$;
    \item $\omega^\mathrm{th}_{\bullet\, n}$: the corresponding emission frequency, i.e., the frequency at which $a_n$ becomes singular under the conditions $G = G^\mathrm{th}_{\bullet\, n}$ and $\omega^g_{\bullet\, n}$.
\end{itemize}

This distinction between $ \omega_n^0 $, $\omega^g_{\bullet\, n}$, and $\omega^\mathrm{th}_{\bullet\, n}$ is conceptually useful to interpret our findings, and it will be referenced throughout the following sections, where we systematically investigate the behavior of the first three modes ($ n = 1, 2, 3 $) in nanoparticles of increasing radii.

Our theoretical framework is general and can be applied to a broad range of materials and geometries. However, in the following, we will focus on a representative system: a silver nanoparticle embedded in an active dielectric host (e.g., water with gain inclusions). This specific configuration allows us to produce concrete results and to visualize the typical behavior of the emission thresholds and frequencies that will be presented in the following sections.

\section*{Results}
\subsection*{Gain Thresholds and Singular Frequencies}

In the previous sections, we introduced a numerical protocol to calculate the threshold gain $ G^\mathrm{th}_{\bullet\, n} $ and the corresponding emission frequency $ \omega^\mathrm{th}_{\bullet\, n} $ for each mode $ n $, under the optimal condition where the gain center frequency $ \omega_g $ is tuned to $ \omega^g_{\bullet\, n} $, minimizing the required gain. Here we show how these quantities depend on the particle radius $ r $, focusing on the first three resonance modes ($ n = 1, 2, 3 $) in the case study of a silver nanoparticle embedded in water ($ \varepsilon_b = 1.7689 $).

For the dielectric function of silver $ \varepsilon_m(\omega) $, we employ two complementary models: (i) the Drude model (see Eq.~\ref{eq:drude} in the Methods), which offers theoretical transparency and highlights general trends, and (ii) the spline interpolation of the experimentally measured dataset by Johnson and Christy~\cite{Johnson:1972}, which enables more realistic comparisons with experimental systems.

\begin{figure}[ht]
\centering
\includegraphics[width=0.8\textwidth]{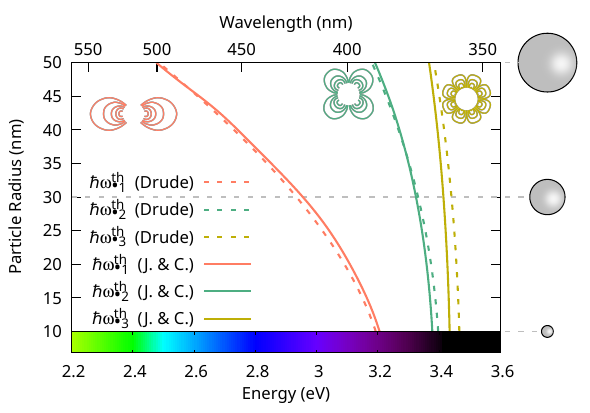}
\caption{Emission frequency for dipolar (orange), quadrupolar (green), and octupolar (yellow) modes as a function of particle size, computed using the Drude model (dashed lines) and the Johnson and Christy experimental data (solid lines).}
\label{fg:reso}
\end{figure}
Figure~\ref{fg:reso} shows the spectral positions of the emission frequencies $ \hbar\omega^\mathrm{th}_{\bullet\, 1} $, $ \hbar\omega^\mathrm{th}_{\bullet\, 2} $, and $ \hbar\omega^\mathrm{th}_{\bullet\, 3} $, corresponding to the dipolar, quadrupolar, and octupolar modes, respectively, as the nanoparticle radius increases from 10 to 50~nm. Despite quantitative differences between the Drude (dashed lines) and Johnson–Christy (solid lines) models, the overall trend is the same: all three emission frequencies redshift with increasing particle size. This shift is most pronounced for the dipolar mode, which moves from approximately 3.2~eV to 2.5~eV. The higher-order modes remain within the 3.2–3.5~eV range.

\begin{figure}[ht]
\centering
\includegraphics[width=0.8\textwidth]{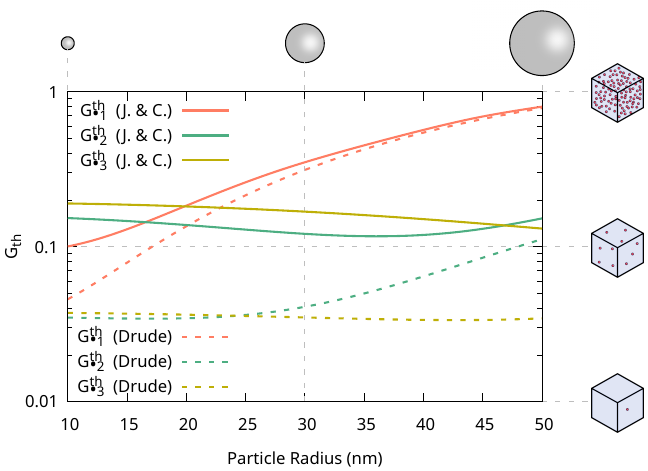}
\caption{Threshold gain for dipolar (orange), quadrupolar (green), and octupolar (yellow) modes as a function of particle size, computed using the Drude model (dashed lines) and the interpolation of the Johnson and Christy experimental data (solid lines).}
\label{fg:thre}
\end{figure}
The corresponding threshold gain values $ G^\mathrm{th}_{\bullet\, 1} $, $ G^\mathrm{th}_{\bullet\, 2} $, and $ G^\mathrm{th}_{\bullet\, 3} $ are shown in Fig.~\ref{fg:thre}. In the Drude model, all three modes begin around $ G \sim 0.04 $ at 10~nm, but diverge as the radius increases: the dipolar threshold rises by nearly an order of magnitude, while the higher-order modes remain flat or even decrease slightly. In the Johnson and Christy model, the dipolar mode initially has the lowest threshold, but surpasses the others for $ r > 20 $~nm. The quadrupolar threshold decreases up to $ r \sim 40 $~nm and then rises, whereas the octupolar threshold decreases consistently across the full range.

These contrasting trends arise from the differing frequency-dependent absorption losses characterizing the two models. A detailed analysis of this effect is deferred to the Discussion section (see Fig.~\ref{fg:imeG}). Highlighting this effect is crucial, as it may otherwise lead to misinterpretations of experimental observations. In particular, the apparent inconsistencies in the gain thresholds required for different modes and sizes may be fully explained once the spectral shape of $ \varepsilon_m''(\omega) $ is taken into account.

The observation that higher-order modes require lower gain thresholds to reach the emission condition is consistent with prior theoretical and numerical work. Several studies have reported that dark or higher-order plasmonic modes—due to their reduced radiative losses and higher quality factors—are more easily driven to emission than their dipolar counterparts. This trend has been established through FDTD and FEM simulations in nanostructures such as core–shell particles~\cite{Yu:2016}, disk-ring spasers~\cite{Huo:2017}, and nanorods~\cite{Liu:2011a}, and confirmed experimentally in nanoparticle arrays supporting both bright and dark modes~\cite{Hakala:2017}. To the best of our knowledge, the present study provides the first systematic characterization of this effect based on a steady-state analytical framework. By identifying the singularities of the complex-valued Mie coefficients~$ a_n(\omega) $ under gain, we enable a direct, mode-resolved comparison of emission thresholds across multiple orders, independent of time-domain simulations or near-field coupling models.

\subsection*{Dye Molecule Density Estimates}

To assess the experimental feasibility of the gain thresholds reported in this work, we estimated the required density of active dye molecules using the relation previously derived in Ref.~\cite{Caicedo:2022}:
\begin{equation}\label{eq:enne}
n_\mathrm{th}=\frac{3\hbar\varepsilon_0\Delta}{2\mu^2 \tilde{N}}G^\mathrm{th},
\end{equation}
This relation links the macroscopic gain parameter $G$ to the microscopic properties of the gain medium. Assuming a fully pumped nanoparticle (i.e., complete population inversion, $\tilde{N} = 1$), a typical emission linewidth of $\hbar\Delta = 0.15$~eV, and a transition dipole moment $\mu$ in the range of $5$–$15$~Debye—consistent with commonly used dyes such as Rhodamine 6G or DCM—we can evaluate the molecular densities and corresponding molar concentrations required to sustain the gain values necessary for emission.

For small nanoparticles (e.g., $r = 10$~nm), all three modes can be driven to emission with relatively modest gain thresholds ($G^\mathrm{th}_{\bullet\, n} \sim 0.1$), resulting in dye densities below $0.2$~nm$^{-3}$ even for low dipole moments ($\mu \sim 5$~Debye). These concentrations are well within the experimentally accessible range for solutions.

For larger nanoparticles (e.g., $r = 50$~nm), the situation becomes more demanding for the dipolar mode, which may require gain values as high as $G^\mathrm{th}_{\bullet\, 1} = 0.8$. This threshold can still be achieved with practical dye concentrations if the gain medium has a sufficiently strong transition dipole moment: for instance, with $\mu = 15$~Debye, the required dye density is approximately $0.1$~nm$^{-3}$, corresponding to about $0.17$~mol/L—well within feasible experimental limits. However, for weaker gain media (e.g., $\mu = 5$~Debye), the required density rises to nearly $0.9$~nm$^{-3}$, approaching or exceeding the close-packing limit~\cite{wiki:cpes}. In contrast, the higher-order modes ($n = 2, 3$) in the same large particles reach emission at significantly lower gain thresholds ($G^\mathrm{th}_{\bullet\, n} \sim 0.15$), and can be driven to emission even with low-efficiency dyes, requiring concentrations well below the packing limit.

\subsection*{Mie Coefficient Spectra}

In what follows, we focus on the electric-type scattering coefficients $ a_n(\omega) $, which exhibit well-defined resonant behavior associated with plasmonic modes in metallic nanoparticles. The corresponding magnetic-type coefficients $ b_n(\omega) $ remain nonresonant and negligible in magnitude across the studied configurations, and are therefore omitted from the analysis (see Methods for details). We present the spectra of the first three Mie scattering coefficients, $ a_1(\omega) $, $ a_2(\omega) $, and $ a_3(\omega) $, for different levels of gain and particle sizes. As in the previous section, we consider silver nanoparticles in water, and to avoid overloading the figures—since the physical behavior does not change significantly—we focus on results obtained using the experimentally interpolated permittivity from Johnson and Christy, which more closely reflects what might be observed in experiments. The gain spectrum is always centered at the optimal frequency $ \omega_g = \omega^g_{\bullet\, n} $ for each mode, corresponding to the minimal gain condition identified earlier.

All figures follow a consistent structure: each row corresponds to a different resonance mode ($n = 1$, $2$, $3$), while each column shows the spectral behavior at a different gain level, expressed as a fraction or multiple of the corresponding threshold $G^\mathrm{th}_{\bullet\, n}$. The background of each panel displays the absolute value of the imaginary part of the gain medium permittivity, $|\varepsilon''_\mathrm{g}(\omega)|$, as a color gradient matched to the spectral curve’s hue. The shading is globally normalized to the maximum value of $|\varepsilon''_\mathrm{g}|$ across all panels in the row, enabling visual comparisons of spectral bandwidth and gain strength between modes and gain levels.
\begin{figure}[ht]
\centering
\includegraphics[width=0.8\textwidth]{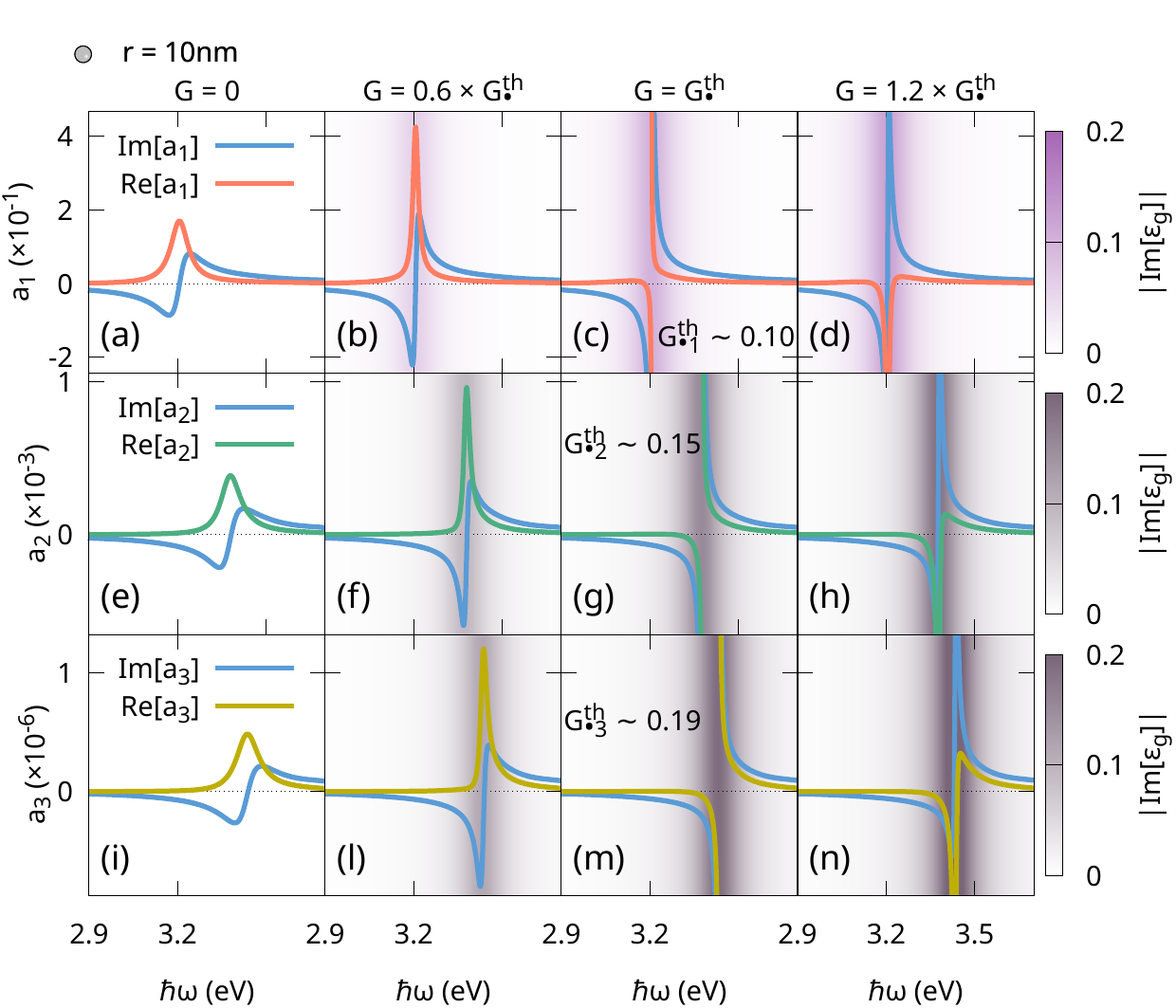}
\caption{Real and imaginary parts of the Mie coefficient spectra for a silver nanoparticle of radius $r=10$~nm for dipolar (row 1), quadrupolar (row 2), and octupolar (row 3) modes. Each column represents a different gain level, from passive to beyond the lasing threshold. The gradient of color in the background represents the absolute value of the imaginary part of the gain medium permittivity.}
\label{fg:s10}
\end{figure}

Figure~\ref{fg:s10} refers to a particle with a radius of $10$~nm. As shown in the first row of the figure (Fig.~\ref{fg:s10}a–d), if the gain emission is centered at the dipolar singular frequency $\hbar\omega^\mathrm{th}_{\bullet\, 1} \sim 3.2$~eV, the first coefficient $a_1(\omega)$ can be driven to a singular behavior when the gain level reaches the corresponding dipolar threshold value $G^\mathrm{th}_{\bullet\, 1} \sim 0.1$ (Fig.~\ref{fg:s10}c). For lower gain levels, the resonance is progressively enhanced (Fig.~\ref{fg:s10}b), while for higher gain levels, the real part of $a_1(\omega)$ becomes negative, indicating that the system enters a nonphysical regime where a steady-state approach is no longer valid~\cite{Recalde:2023}. This behavior mirrors the quasi-static scenario previously analyzed in Refs.~\cite{Veltri:2012,Caicedo:2022}, where the singularity in the polarizability similarly marked the emission threshold. Despite the breakdown of the linear steady-state model beyond this point, the appearance of the singularity remains a reliable criterion for the onset of emission/spasing.

The second (Fig.~\ref{fg:s10}e–h) and third rows (Fig.~\ref{fg:s10}i–n) show that similar behavior occurs for the quadrupolar ($a_2(\omega)$) and octupolar ($a_3(\omega)$) modes, respectively. When the gain frequency is tuned to the corresponding resonance ($\hbar\omega^\mathrm{th}_{\bullet\, 2} \sim 3.38$~eV and $\hbar\omega^\mathrm{th}_{\bullet\, 3} \sim 3.44$~eV) and the appropriate gain thresholds are reached ($G^\mathrm{th}_{\bullet\, 2} \sim 0.15$, $G^\mathrm{th}_{\bullet\, 3} \sim 0.2$), a singular behavior again emerges in the respective Mie coefficients (see Fig.~\ref{fg:s10}g and Fig.~\ref{fg:s10}m).

\begin{figure}[ht]
\centering
\includegraphics[width=0.8\textwidth]{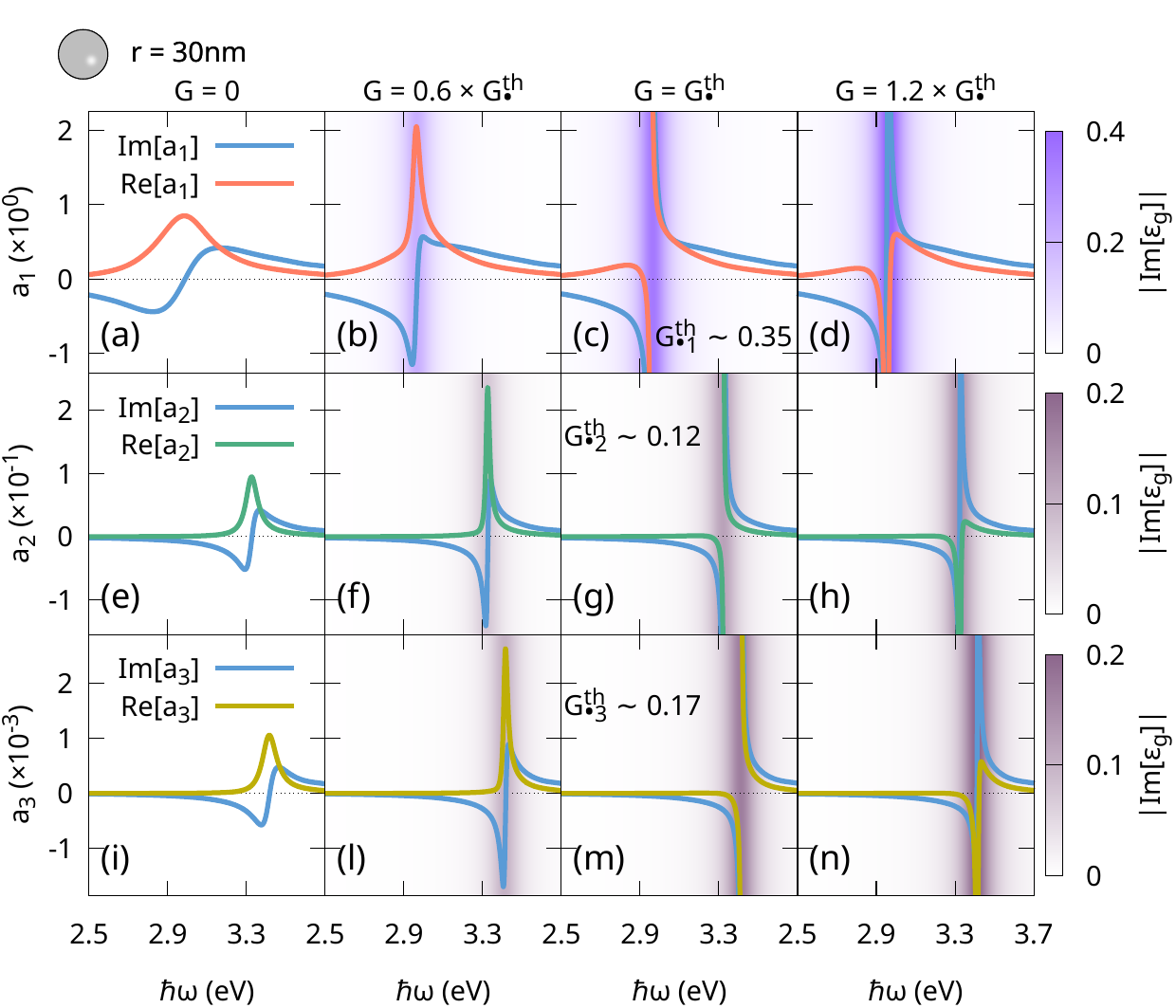}
\caption{Real and imaginary parts of the Mie coefficient spectra for a silver nanoparticle of radius $r=30$~nm for dipolar (row 1), quadrupolar (row 2), and octupolar (row 3) modes. Each column represents a different gain level, from passive to beyond the lasing threshold. The gradient of color represents the absolute value of the imaginary part of the gain medium permittivity.}
\label{fg:s30}
\end{figure}
Figure~\ref{fg:s30} presents the behavior of the Mie coefficients for a nanoparticle with radius $r = 30$~nm. As in the $10$~nm case, all three modes can be driven to emission when the gain is centered at their respective optimal frequencies and reaches the corresponding threshold values (see panels~\ref{fg:s30}c, g, and m). The quadrupolar and octupolar spectra (rows two and three) retain a structure very similar to that observed for the smaller particle, displaying sharp, symmetric features that clearly develop into singularities as gain increases. The dipolar response (row one), however, already begins to differ notably: while the passive spectrum (Fig.~\ref{fg:s30}a) shows just a broader resonance compared to the $10$~nm case, the asymmetry becomes clearly visible at $0.6\,G^\mathrm{th}_{\bullet\, 1}$ (Fig.~\ref{fg:s30}b), where the real part of $a_1(\omega)$ drops sharply on the low-frequency side and decays more gradually on the high-frequency side.

Additionally, the overall amplitude of the Mie coefficients increases significantly with size: the maximum of $|a_1|$ grows by an order of magnitude compared to the smaller particle ($r = 10$~nm), while $|a_2|$ and $|a_3|$ increase by roughly two and three orders of magnitude, respectively. This substantial enhancement highlights the increasing relevance of higher-order modes as the system moves beyond the quasi-static regime.

\begin{figure}[ht]
\centering
\includegraphics[width=0.8\textwidth]{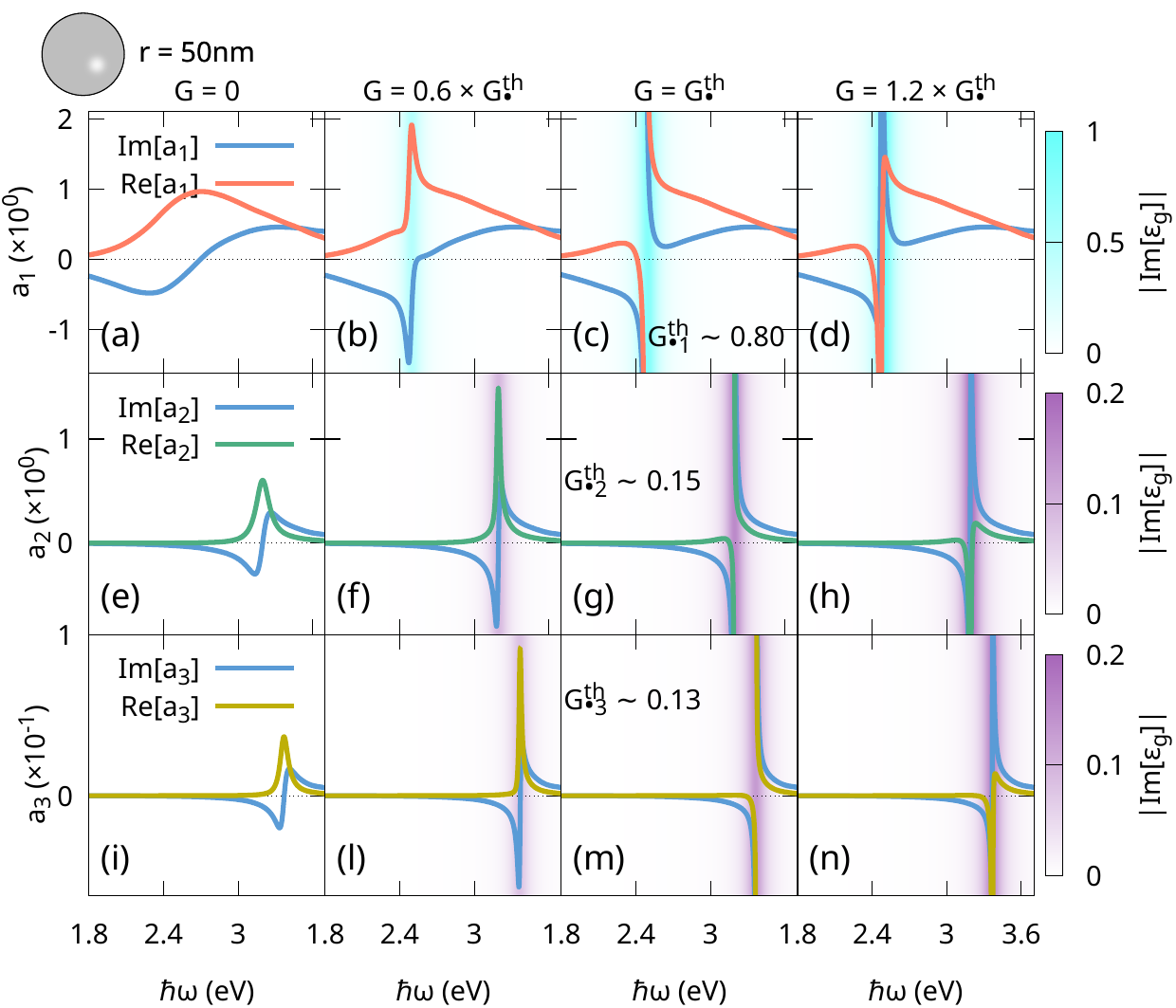}
\caption{Real and imaginary parts of the Mie coefficient spectra for a silver nanoparticle of radius $r=50$~nm for dipolar (row 1), quadrupolar (row 2), and octupolar (row 3) modes. Each column represents a different gain level, from passive to beyond the lasing threshold. The gradient of color represents the absolute value of the imaginary part of the gain medium permittivity.}
\label{fg:s50}
\end{figure}
This trend becomes even more evident in Figure~\ref{fg:s50}, which shows the spectra for a particle with radius $r = 50$~nm. Here, the dipolar resonance is significantly broader and more asymmetric than in the previous two cases (Fig.~\ref{fg:s50}a–d). While the singularity condition is still satisfied when gain reaches $G^\mathrm{th}_{\bullet\, 1}$ (Fig.~\ref{fg:s50}c), the misalignment between the optimal gain center frequency $\hbar\omega^\mathrm{g}_{\bullet\, 1}$ and the actual emission frequency $\hbar\omega^\mathrm{th}_{\bullet\, 1}$ becomes more pronounced—reflecting the breakdown of the narrow-resonance assumption that holds in the quasi-static regime. In contrast, the quadrupolar and octupolar modes maintain narrower and more symmetric profiles, and their optimal and emission frequencies remain closely aligned even at this larger size (see Fig.~\ref{fg:s50}g and m). 

This comparison illustrates a key point: while all modes can be driven to emission across the particle sizes explored, higher-order modes (especially quadrupolar and octupolar) exhibit greater spectral stability and better-defined thresholds, making them potentially more robust candidates for emission in larger plasmonic systems.

\subsection*{Extinction Cross Section}

While the singular behavior of individual Mie coefficients $a_n$ marks the onset of emission for the corresponding mode, it is through the extinction cross section that the collective optical response of the nanoparticle becomes manifest. In particular, the extinction spectrum reflects the combined contributions of multiple modes and their interplay, providing a more experimentally accessible observable. For instance, even when only a single Mie coefficient becomes singular, the total extinction spectrum reveals how this transition influences or dominates the overall response of the system. Conversely, when two or more modes are close to threshold, the extinction cross section allows one to visualize possible modal competition, suppression, or enhancement effects.

In this context, the role of the gain medium becomes central. As described in the Methods section, the host dielectric is modeled by a complex, frequency-dependent permittivity $\varepsilon_\mathrm{g}(\omega)$, given by a Lorentzian profile centered at the gain frequency $\omega_g$ and characterized by a gain strength $G$ and linewidth $\Delta$. The associated refractive index $n_2 = \sqrt{\varepsilon_\mathrm{g}}$ is thus also complex, and so is the wavevector $k = \frac{\omega}{c} n_2 = k' + i k''$. This introduces an intrinsic challenge when computing cross sections, which are typically defined in terms of the real part of $k$.

In particular, when $k'' \ne 0$, standard formulas used for passive media—where the wavevector is purely real—can lead to ill-defined or even complex-valued cross sections. This is because the presence of gain or loss in the host medium violates the assumptions under which conventional definitions of scattering, absorption, and extinction cross sections were derived. In response to this, the literature has proposed various generalized formulations to handle absorbing (or amplifying) host media, distinguishing between \emph{inherent} and \emph{apparent} cross sections~\cite{Mundy1974,Chylek1979,Lebedev1999,Sudiarta2001,Yang2002,Fu2006,Bohren1979,Mishchenko2007,Mishchenko2017,Mishchenko2018,Mishchenko2019}.

Among these, the \emph{apparent extinction cross section}—based on the scattered far-field power normalized by the real part of the host wavevector—is widely used in the Mie theory literature, and yields physically meaningful, real-valued quantities even in the presence of gain or absorption. In this work, we adopt this apparent definition:
\begin{equation}
    C_{\mathrm{ext}} = \frac{2 \pi}{k'} \mathrm{Re} \left\{ \sum^{\infty}_{n=1} \frac{1}{k} (2n+1) (a_n + b_n) \right\},
\end{equation}
where $ a_n $ and $ b_n $ are the electric and magnetic Mie scattering coefficients, respectively.

This approach allows us to study how the emergence of singularities in specific $a_n$ affects the extinction spectrum, and to assess how multiple modes contribute collectively to the optical response under different gain conditions.
To construct the plots in this section, we computed the extinction efficiency spectrum $ C_{\mathrm{ext}}/(\pi r^2) $ using up to 16 multipolar modes ($ n = 1 $ to $ 16 $), which exceeds the value recommended by the Wiscombe criterion~\cite{Wiscombe1980} across our entire size and wavelength range. In the most demanding configuration, this criterion yields $ N_\mathrm{max} \approx 6.5 $, confirming that our choice ensures complete convergence of the Mie series even for the largest particles considered, where higher-order contributions become increasingly relevant. 

Each figure is arranged such that rows correspond to different gain configurations—specifically, with the gain spectrum centered at the emission frequency $ \omega^\mathrm{th}_{\bullet\,1} $, $ \omega^\mathrm{th}_{\bullet\,2} $, and $ \omega^\mathrm{th}_{\bullet\,3} $ for the dipolar, quadrupolar, and octupolar modes, respectively. Meanwhile, columns display the extinction spectrum for increasing levels of gain, from passive ($ G = 0 $) to above the lasing threshold.

Panels (a), (e), and (i) in each figure show identical extinction spectra computed for the passive case $ G = 0 $, serving as reference baselines for comparison with gain-modified spectra in the same row. While they are formally identical, they are included at the beginning of each row to aid in visualizing the effect of mode-selective gain activation. The extinction efficiency curves for the passive system were independently validated against the results from the online Mie scattering tool provided by Saviot~\cite{saviot_mie}, ensuring the consistency of our implementation.

\begin{figure}[ht]
\centering
\includegraphics[width=0.8\textwidth]{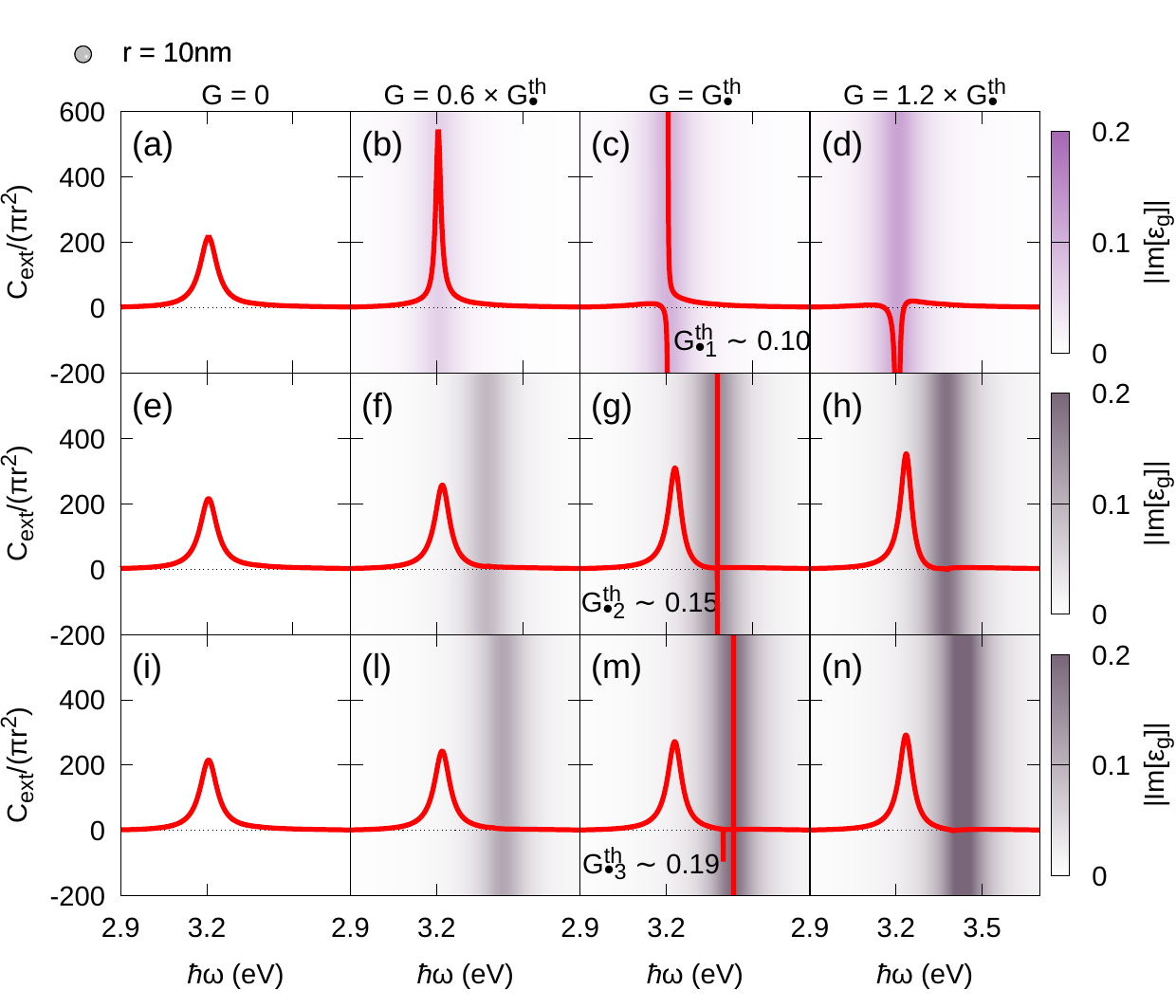}
\caption{
Extinction cross section spectra for a silver nanoparticle of radius $r = 10$~nm. Each row corresponds to a different gain alignment: the gain spectrum is centered at $\omega^\mathrm{th}_{\bullet\, 1}$ (row 1), $\omega^\mathrm{th}_{\bullet\, 2}$ (row 2), and $\omega^\mathrm{th}_{\bullet\, 3}$ (row 3), respectively. Each column shows a different gain level, from passive to beyond the lasing threshold. The background gradient illustrates the absolute value of the imaginary part of the gain medium permittivity $|\varepsilon''_\mathrm{g}(\omega)|$.}
\label{fg:cs10}
\end{figure}
Figure~\ref{fg:cs10} presents the extinction efficiency spectra for a silver nanoparticle of radius $r = 10$~nm under increasing levels of gain, with each row corresponding to a different choice of gain center frequency: the dipolar ($\omega^\mathrm{th}_{\bullet\,1}$), quadrupolar ($\omega^\mathrm{th}_{\bullet\,2}$), and octupolar ($\omega^\mathrm{th}_{\bullet\,3}$) resonances, respectively.

The first row (panels~\ref{fg:cs10}a--d), where the gain is centered on the dipolar mode, reproduces the behavior expected from the $a_1(\omega)$ spectra computed under the same gain conditions. The zero-gain extinction curve (panel~a) features a narrow and well-defined resonance. As gain increases to $0.6\,G^\mathrm{th}_{\bullet\,1}$ (panel~b), the dipolar peak becomes both narrower and more intense. At threshold (panel~c), the singular behavior manifests clearly, and for gain exceeding the threshold (panel~d), the extinction becomes negative in the resonance region—a nonphysical result indicating the breakdown of the steady-state linear model. Notably, in this configuration, the contributions of higher-order modes are negligible, and the overall response is effectively quasi-static.

The second row (panels~\ref{fg:cs10}e--h), with gain centered on the quadrupolar resonance, reveals a more intricate interplay. In the sub-threshold regime (panel~f), the quadrupolar peak remains largely invisible compared to the dominant dipolar contribution. Only at threshold (panel~g) does the quadrupolar resonance emerge with sufficient intensity to be distinguished from the dipolar background. For gain above threshold (panel~h), the quadrupolar mode begins to display a shallow negative dip—a nonphysical artifact resulting from the breakdown of the linear steady-state model. This highlights the limitations of the formalism in describing post-threshold behavior and should not be interpreted as actual suppression of the mode. In fact, as demonstrated in the dynamical model of Ref.~\cite{Aradian:2024}, additional gain beyond threshold leads to increased emission, not diminished response.

The third row (panels~\ref{fg:cs10}i–n), where gain is centered on the octupolar mode, displays qualitatively similar behavior. The octupolar resonance becomes prominent only at threshold (panel~m); in the same panel, the quadrupolar peak appears as a weak, nonphysical dip. This indicates that, within the fixed emission bandwidth of the gain medium ($\Delta = 0.15$~eV), the quadrupolar mode is preferentially amplified over the octupolar one. As a result, even when gain is centered at the octupolar resonance, the system may reach the quadrupolar emission threshold first—at an intermediate gain level between panels~l and~m—due to its more favorable coupling conditions.

\begin{figure}[ht]
\centering
\includegraphics[width=0.8\textwidth]{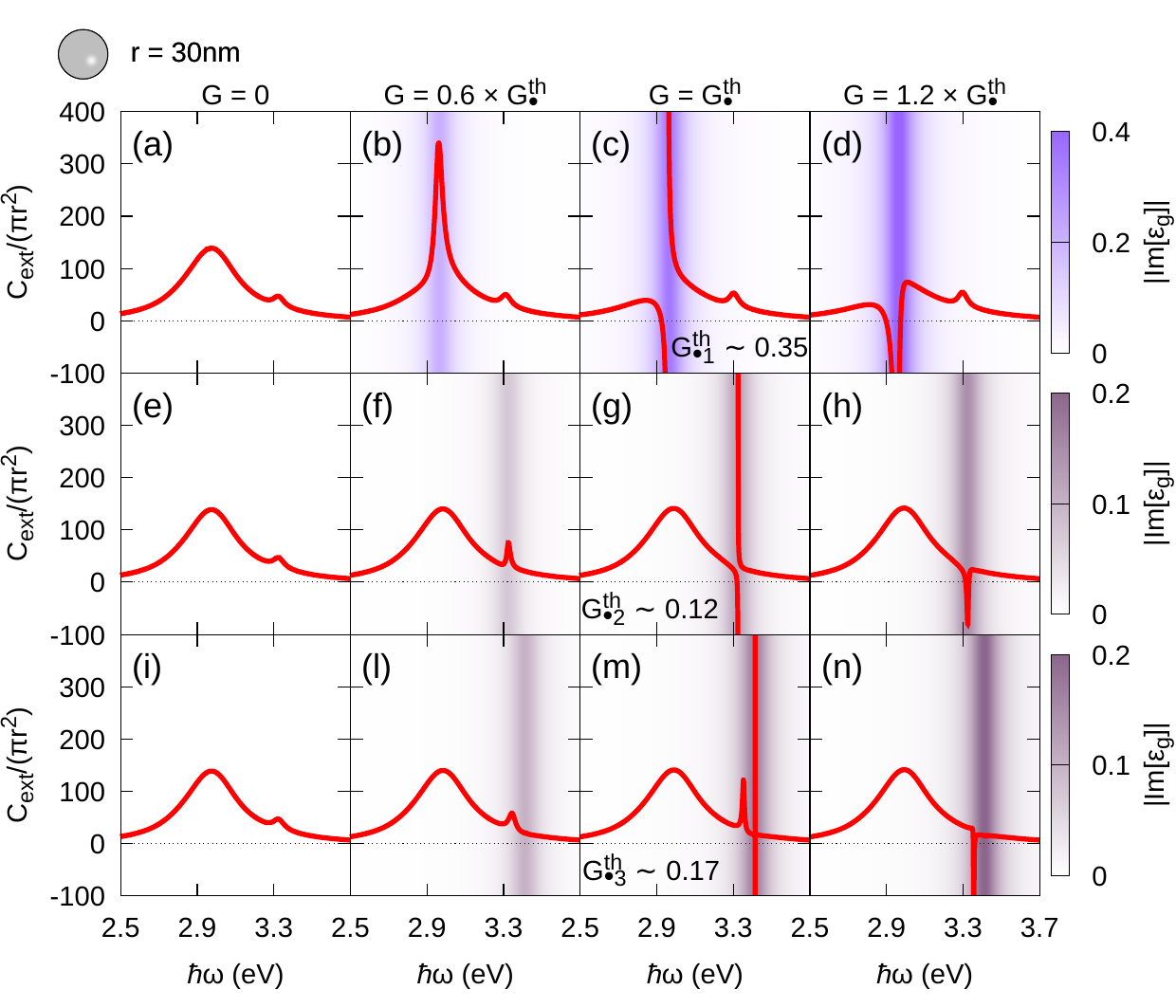}
\caption{
Extinction cross section spectra for a silver nanoparticle of radius $r = 30$~nm. Each row corresponds to a different gain alignment: the gain spectrum is centered at $\omega^\mathrm{th}_{\bullet\, 1}$ (row 1), $\omega^\mathrm{th}_{\bullet\, 2}$ (row 2), and $\omega^\mathrm{th}_{\bullet\, 3}$ (row 3), respectively. Each column shows a different gain level, from passive to beyond the lasing threshold. The background gradient illustrates the absolute value of the imaginary part of the gain medium permittivity $|\varepsilon''_\mathrm{g}(\omega)|$.}
\label{fg:cs30}
\end{figure}
In the case of the $30$~nm radius nanoparticle (Fig.~\ref{fg:cs30}), the G = 0 column (panels~a, e, i) already reveals a key difference compared to the $10$~nm case: the dipolar resonance is now broader, and a secondary peak appears at higher energy, corresponding to the contribution of the quadrupolar mode. In the first row, where the gain is centered on the dipolar emission frequency $\hbar\omega^\mathrm{th}_{\bullet\,1}$, it is possible to drive the dipolar mode to emission without significantly affecting the secondary peak associated with the quadrupolar mode. In the second row, the gain is centered on $\hbar\omega^\mathrm{th}_{\bullet\,2}$, and it is now the secondary peak that is driven to emission, while the rest of the spectrum remains mostly unchanged.

The third row illustrates that, at this particle size, the octupolar mode can indeed be driven to emission before the quadrupolar one, as their spectral features are now well separated. However, once the gain reaches the threshold for the octupolar mode (panel~\ref{fg:cs30}m), the quadrupolar peak becomes strongly amplified. In the nonphysical regime above threshold (panel~\ref{fg:cs30}n), it is the quadrupolar mode that dominates the spectral response with a pronounced negative dip, while the octupolar feature becomes less visible. Although this regime lies outside the validity of the steady-state model, it carries a practical implication: if the gain is finely tuned to the threshold of the octupolar mode, it is possible to selectively activate this higher-order mode. However, if excessive gain is introduced—even if centered at $\hbar\omega^\mathrm{th}_{\bullet\,3}$—the enhanced amplification of the quadrupolar mode may overshadow the octupolar emission, reducing mode selectivity.

\begin{figure}[ht]
\centering
\includegraphics[width=0.8\textwidth]{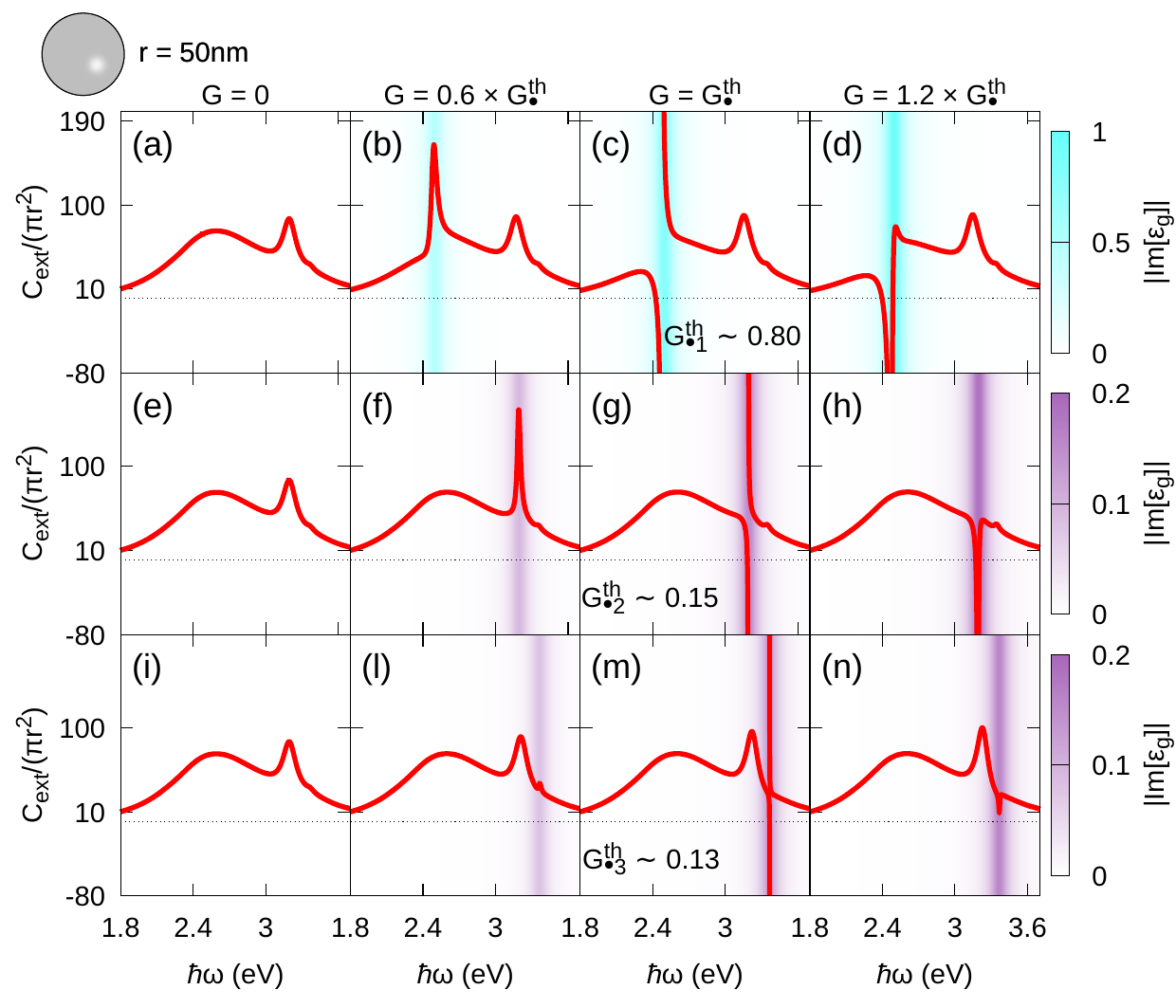}
\caption{
Extinction cross section spectra for a silver nanoparticle of radius $r = 50$~nm. Each row corresponds to a different gain alignment: the gain spectrum is centered at $\omega^\mathrm{th}_{\bullet\, 1}$ (row 1), $\omega^\mathrm{th}_{\bullet\, 2}$ (row 2), and $\omega^\mathrm{th}_{\bullet\, 3}$ (row 3), respectively. Each column shows a different gain level, from passive to beyond the lasing threshold. The background gradient illustrates the absolute value of the imaginary part of the gain medium permittivity $|\varepsilon''_\mathrm{g}(\omega)|$.}
\label{fg:cs50}
\end{figure}
For the $50$~nm radius nanoparticle (Fig.~\ref{fg:cs50}), the zero-gain spectrum (column~a, e, i) reveals prominent multipolar contributions: in addition to the broader and asymmetric dipolar resonance, a distinct quadrupolar peak and a discernible octupolar bump are already present. These features confirm that the system lies well beyond the quasi-static regime, with higher-order modes contributing significantly to the optical response even in the absence of gain.

What sets this size regime apart from the smaller cases is that the spectral separation between the modes is now sufficient to allow clean and selective excitation of each individual resonance. In the first row, the dipolar mode is driven to emission with little interference from the higher modes. In the second row, the quadrupolar resonance is selectively amplified, leaving the dipolar and octupolar contributions mostly unaffected. Most notably, in the third row, the octupolar mode can finally be driven to emission without inducing significant amplification—or premature emission—in the quadrupolar mode. This is in stark contrast to the $30$~nm case, where the two modes were still spectrally entangled, and the quadrupolar mode became dominant even when the gain was tuned to favor the octupolar one.

This ability to selectively excite higher-order modes at larger sizes opens promising avenues for mode-specific emission control in plasmonic systems, and suggests that properly engineered gain profiles and particle sizes can be used to access targeted plasmonic resonances even in complex, multimode environments.

\section*{Methods}

As discussed in Refs.~\cite{Veltri:2012,Veltri:2016,Caicedo:2022,Aradian:2024}, the onset of emission in gain-enhanced metal nanoparticles can be identified with the appearance of a singularity—i.e., a complex zero of the denominator—in either the quasi-static polarizability or, beyond the QS regime, in the Mie scattering coefficients $ a_n $. In particular, it has been shown that the singularity of $ a_n(\omega) $ reliably marks the threshold gain $ G^\mathrm{th}_n $ and the corresponding emission frequency $ \omega^\mathrm{th}_n $ for each multipolar mode $ n $~\cite{Recalde:2023}. 

In this work, we apply this criterion systematically across dipolar, quadrupolar, and octupolar modes, and implement a dedicated numerical scheme to identify the singularities of $ a_n(\omega) $ as a function of both gain and frequency. The goal is to extract, for each mode and particle size, the optimal gain center frequency $ \omega^g_{\bullet\, n} $ that minimizes the gain required to reach the singularity.

The Mie scattering coefficients are calculated using the standard optical approximation expressions~\cite{bohren98}:
\begin{equation}
    \label{eq:an}
    a_n = \frac{m^2 j_n(mx) [x j_n(x)]' - j_n(x) [mx j_n(mx)]'}{m^2 j_n(mx) [x h_n^{(1)}(x)]' - h_n^{(1)}(x) [mx j_n(mx)]'}, 
\end{equation}
\begin{equation}
    \label{eq:bn}
    b_n = \frac{j_n(mx) [x j_n(x)]' - j_n(x) [mx j_n(mx)]'}{j_n(mx) [x h_n^{(1)}(x)]' - h_n^{(1)}(x) [mx j_n(mx)]'},
\end{equation}
where $ m = n_1/n_2 $ is the relative refractive index, $ x = ka $ is the size parameter, and $ j_n $, $ h_n^{(1)} $ are spherical Bessel and Hankel functions, respectively.

Although the full Mie expansion includes both electric- and magnetic-type coefficients ($ a_n $ and $ b_n $), we focus our threshold analysis exclusively on the $ a_n $ modes. These dominate the optical response in the plasmonic regime, exhibiting well-defined resonances and complex poles associated with collective oscillations of conduction electrons. In contrast, $ b_n $ coefficients remain nonresonant and are typically two or more orders of magnitude smaller across the studied spectral and size ranges. While large gain values can induce peaks in $ b_n $, the required thresholds ($ G > 10 $) are well beyond physical values and have no observable influence on the singularities of $ a_n $. Nevertheless, $ b_n $ is retained in the computation of total extinction cross sections, as reported in the section dedicated to the Extinction Cross Sections.

It is worth noting that in high-index dielectric nanoparticles such as silicon or germanium, the magnetic-type Mie coefficients $ b_n $ can become resonant and comparable in strength to $ a_n $, leading to fundamentally different modal dynamics. In such systems, gain-assisted directional scattering and modal interference effects—such as those discussed in Ref.~\cite{Olmos-Trigo:2020}—may become accessible through the selective amplification of magnetic modes. Extending the present threshold analysis to those materials represents a promising direction for future work.

The index of refraction $n_2$ of the surrounding gain medium is derived from its complex permittivity $\varepsilon_\mathrm{g}(\omega)$, modeled using a standard Lorentzian profile:
\begin{equation}
    \varepsilon_\mathrm{g}(\omega) = \varepsilon_\mathrm{b} + \frac{\varepsilon_0 G \Delta}{2(\omega - \omega_\mathrm{g}) + i\Delta}, \label{eq:epsgN}
\end{equation}
where $G$ controls the gain strength, $\omega_\mathrm{g}$ is the gain center frequency, $\Delta$ is the full linewidth, and $\varepsilon_\mathrm{b}$ is the background permittivity of the host (e.g., water, with $\varepsilon_\mathrm{b} = 1.7689$).

For the nanoparticle permittivity $\varepsilon_\mathrm{m}(\omega)$, two models are employed. The first is the Drude model:
\begin{equation}
    \varepsilon_\mathrm{m}(\omega) = \varepsilon_\infty - \frac{\varepsilon_0 \omega_\mathrm{p}^2}{\omega(\omega + 2i\gamma)}, \label{eq:drude}
\end{equation}
with $\varepsilon_\infty = 5.3$, $\omega_\mathrm{p} = 9.6$~eV, and $\gamma = 0.0228$~eV. The second model uses interpolated experimental data from Johnson and Christy~\cite{Johnson:1972}, implemented via cubic spline routines from the GNU Scientific Library (GSL).

Finding an analytical expression for the complex zeros of the denominator in equation~\ref{eq:an} is a non-trivial task. In this work, we developed a robust multi-step procedure that identifies the singularity in the Mie scattering coefficient $ a_n(\omega) $ under optimal gain conditions to determine the mode-specific threshold gain $ G^\mathrm{th}_{\bullet\, n} $ and the corresponding emission frequency $ \omega^\mathrm{th}_{\bullet\, n} $. The method proceeds as follows.

First, the zero-gain resonance frequency $ \omega^0_n $—defined as the frequency at which $ |a_n(\omega)|^2 $ is maximal in the absence of gain—is computed by scanning for the real-frequency maximum of the Mie coefficient modulus. This serves as an initial estimate for the optimal gain center frequency $ \omega^g_{\bullet\, n} $, which we then refine.

To determine the actual emission frequency and the minimal gain required to reach the singular regime, we scan a small spectral interval around $ \omega^0_n $ and minimize the threshold gain over this range. For each trial value of the gain center frequency $ \omega_g $, we compute the gain $ G^\mathrm{th}_n(\omega_g) $ required to induce a singularity in $ a_n(\omega) $. This is done by a two-stage procedure: a coarse-to-fine bisection search combined with a Newton-Raphson refinement in two dimensions—frequency and gain—targeting the complex zero of the denominator of the Mie coefficient.

The bisection search identifies a preliminary estimate of $ G^\mathrm{th}_n $ for a given $ \omega_g $, defined as the smallest gain for which the real part of $ a_n $ becomes negative at resonance. This estimate is then refined by solving for the point where both the real and imaginary parts of the Mie denominator vanish simultaneously. This is achieved by a Newton-Raphson algorithm applied to the determinant of the scattering matrix, using finite-difference approximations to construct the Jacobian in the two-dimensional parameter space $ (\omega, G) $.

The procedure returns four quantities:
\begin{itemize}
  \item $ \omega^g_{\bullet\, n} $: the optimal gain center frequency minimizing the threshold gain;
  \item $ G^\mathrm{th}_{\bullet\, n} $: the minimal threshold gain required to reach the singularity;
  \item $ \omega^\mathrm{th}_{\bullet\, n} $: the corresponding emission frequency at which the singularity occurs;
  \item $ \omega^0_n $: the zero-gain resonance frequency.
\end{itemize}

This method ensures stable and accurate identification of the emission conditions for each mode $ n $, even beyond the quasi-static limit. It also guarantees that the gain is applied under optimal spectral tuning, which is essential for comparing emission thresholds across different modes and particle sizes.

\begin{figure}[ht]
\centering
\includegraphics[width=0.8\textwidth]{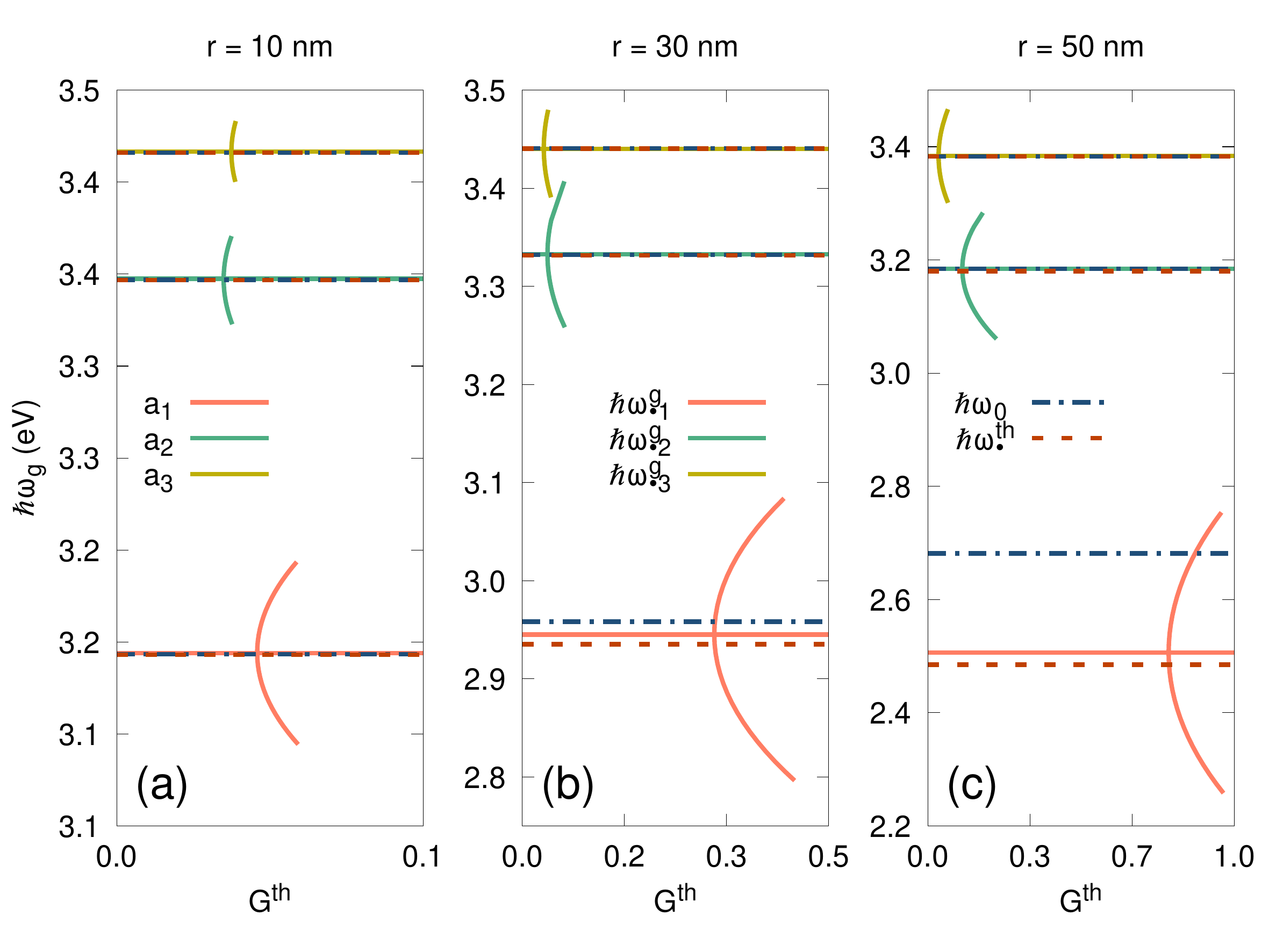}
\caption{
Threshold gain $ G^\mathrm{th} $ versus center frequency $\hbar \omega_g$ of the gain medium for the first three modes: $a_1$ (orange), $a_2$ (green), and $a_3$ (yellow).
Vertical solid lines of the corresponding colors mark the optimal conditions ($ G^\mathrm{th} = G^{\mathrm{th}}_{\bullet}$).
Dashed blue lines show the passive plasmon frequency $\hbar \omega_0$, and dashed brown lines the emission frequency $\hbar \omega^{\mathrm{th}}_{\bullet}$.
Panels: (a) $r=10$~nm; (b) $r=30$~nm; (c) $r=50$~nm.
}
\label{fg:fcr}
\end{figure}
Figure~\ref{fg:fcr} illustrates the full dependence of the threshold gain $ G^\mathrm{th} $ on the center frequency of the gain medium $ \hbar \omega_g $, for the first three modes $ a_1 $ (orange), $ a_2 $ (green), and $ a_3 $ (yellow), across three different particle sizes: $r = 10$~nm, $30$~nm, and $50$~nm. This figure makes visually explicit the point summarized in Table~\ref{tb:thresholds}: for each mode, there exists a well-defined optimal gain center frequency $ \omega^g_{\bullet\, n} $ that minimizes the gain required to induce emission. These optimal points are marked by vertical solid lines in the figure, while the dashed blue and brown lines indicate the zero-gain resonance frequency $ \omega^0_n $ and the emission frequency $ \omega^\mathrm{th}_{\bullet\, n} $, respectively.

While the minimal gain tends to occur close to the passive resonance $ \omega^0_n $, it is evident—particularly for the dipolar mode in particles of radius $ r = 30 $~nm and $ 50 $~nm—that the three characteristic frequencies $ \omega^0_n $, $ \omega^g_{\bullet\, n} $, and $ \omega^\mathrm{th}_{\bullet\, n} $ do not coincide. This discrepancy becomes more pronounced as the dipolar resonance broadens and becomes more asymmetric with increasing particle size, a feature already discussed in Figure~\ref{fg:s30} and Figure~\ref{fg:s50}. Conversely, for small particles (e.g., $r = 10$~nm) and for the higher-order modes ($n = 2, 3$), the three frequencies remain closely aligned.

We do not yet have an analytical explanation for the observed decoupling between the optimal gain center and the emission frequency in certain cases, but we suspect it is related to the spectral mismatch between the mode linewidth and the gain profile. In our simulations, we consistently use a gain linewidth of $\Delta = 0.15$~eV; when the modal resonance becomes significantly broader than this value—particularly for the dipolar mode in larger particles—the spectral overlap condition that minimizes $ G^\mathrm{th} $ may shift, causing a divergence between $ \omega^g_{\bullet\, n} $ and $ \omega^\mathrm{th}_{\bullet\, n} $.

It is also worth noting that for each mode, there exists a specific gain center frequency $ \omega_g $ such that the emission occurs at the same frequency ($ \omega_g = \omega^\mathrm{th} $). This special case corresponds to what could be called “in-phase emission”. However, this condition does not generally coincide with the optimal one and requires a higher threshold gain to be fulfilled.

\section*{Discussion}

In this work, we developed a protocol to determine the threshold gain $ G^\mathrm{th}_{\bullet\, n} $ and corresponding emission frequency $ \omega^\mathrm{th}_{\bullet\, n} $ for each multipolar resonance $ n $, based on identifying singularities in the Mie scattering coefficients under gain. This mode-resolved approach enables direct evaluation of emission conditions without relying on time-domain simulations or coupled-mode approximations.

Our analysis shows that $ G^\mathrm{th}_{\bullet\, n} $ is highly sensitive to the dielectric model used for silver. Comparing the Drude and Johnson–Christy permittivities reveals qualitatively different trends across particle sizes and resonance orders, stemming from their distinct frequency-dependent absorption losses.

\begin{figure}[ht]
\centering
\includegraphics[width=0.8\textwidth]{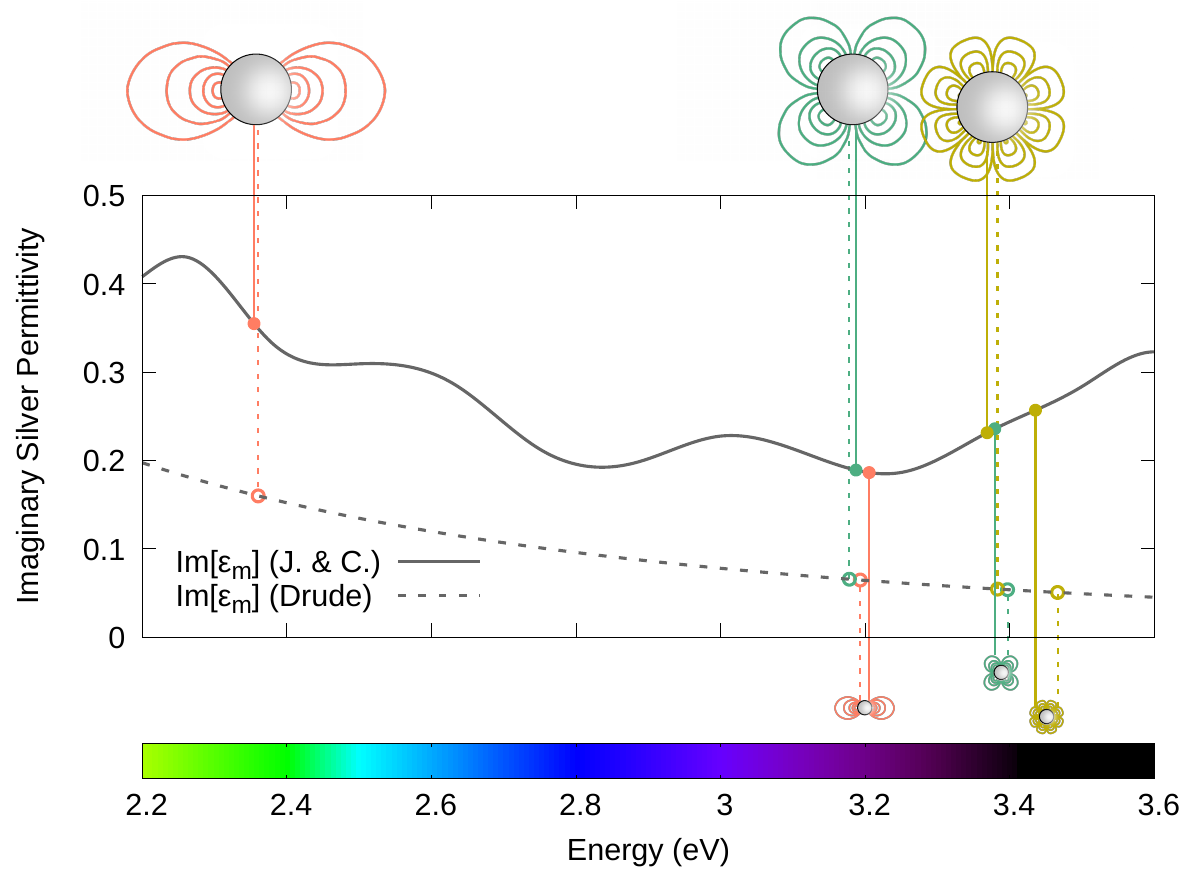}
\caption{Imaginary part of the silver permittivity $ \varepsilon_m''(\omega) $. The dashed line corresponds to the Drude model; the solid line to the Johnson and Christy dataset~\cite{Johnson:1972}. Vertical markers indicate the emission frequencies $ \hbar\omega^\mathrm{th}_{\bullet\,1} $, $ \hbar\omega^\mathrm{th}_{\bullet\,2} $, and $ \hbar\omega^\mathrm{th}_{\bullet\,3} $ for radii of 10~nm and 50~nm.}
\label{fg:imeG}
\end{figure}

Figure~\ref{fg:imeG} clarifies this effect. For small particles ($ r = 10 $~nm), the dipolar emission frequency falls in a low-loss region of the Johnson–Christy dataset, whereas higher-order modes lie in more absorptive regions. As particle size increases, all resonances redshift, with higher modes reaching lower-loss spectral zones—leading to decreased thresholds. In contrast, the Drude model's monotonic decay in $ \varepsilon_m'' $ with frequency results in a uniform increase in thresholds with size, capturing general size trends but missing mode-specific reversals.

It is crucial to highlight this aspect, as it can otherwise lead to misinterpretations of experimental results. Apparent inconsistencies in gain thresholds across modes and sizes are readily explained by the spectral structure of $ \varepsilon_m''(\omega) $.

More generally, these results confirm that $ G^\mathrm{th}_{\bullet\, n} $ trends are intimately tied to the metallic permittivity. In an idealized plasmonic system—well described by the Drude model—higher-order modes exhibit lower thresholds due to reduced radiative losses and higher quality factors. However, when realistic losses from interband transitions are included, these trends may be masked or inverted. Our mode-specific method enables disentangling such effects, providing insight into how size, mode order, and material dispersion interact in gain-assisted emission.

The estimated dye concentrations required to sustain these thresholds remain experimentally feasible, as discussed in the subsection \emph{Dye Molecule Density Estimates}.

\begin{table*}[ht]
\small
\caption{
Summary of spectral and gain parameters for the first three multipolar modes ($a_1$, $a_2$, $a_3$) of silver nanoparticles with radii $r = 10$, $30$, and $50$ nm. Reported values include the zero-gain resonance frequency $\hbar\omega^0$, the optimal gain center frequency $\hbar\omega^\mathrm{g}{\bullet}$, the emission frequency $\hbar\omega^\mathrm{th}{\bullet}$, and the corresponding threshold gain $G^\mathrm{th}_{\bullet}$ (dimensionless), all rounded to three decimal places.
}
\label{tb:thresholds}
\begin{tabular*}{\textwidth}{@{\extracolsep{\fill}}|l|llll|llll|llll|}
\hline
& \multicolumn{4}{c|}{$r=10$~nm} & \multicolumn{4}{c|}{$r=30$~nm} & \multicolumn{4}{c|}{$r=50$~nm} \\
& $\hbar\omega^0$ & $\hbar\omega^\mathrm{g}_{\bullet}$ & $\hbar\omega^\mathrm{th}_{\bullet}$ & $G^\mathrm{th}_{\bullet}$
& $\hbar\omega^0$ & $\hbar\omega^\mathrm{g}_{\bullet}$ & $\hbar\omega^\mathrm{th}_{\bullet}$ & $G^\mathrm{th}_{\bullet}$
& $\hbar\omega^0$ & $\hbar\omega^\mathrm{g}_{\bullet}$ & $\hbar\omega^\mathrm{th}_{\bullet}$ & $G^\mathrm{th}_{\bullet}$ \\
\hline
$a_1$ & 3.193 & 3.194 & 3.193 & 0.046 & 2.958 & 2.945 & 2.935 & 0.313 & 2.682 & 2.506 & 2.485 & 0.786 \\
$a_2$ & 3.397 & 3.398 & 3.397 & 0.035 & 3.332 & 3.333 & 3.332 & 0.041 & 3.185 & 3.184 & 3.180 & 0.112 \\
$a_3$ & 3.466 & 3.467 & 3.466 & 0.037 & 3.441 & 3.440 & 3.440 & 0.035 & 3.383 & 3.384 & 3.383 & 0.035 \\
\hline
\end{tabular*}
\end{table*}

Table~\ref{tb:thresholds} summarizes the extracted emission parameters. Two key points emerge: First, the frequency offset between $\omega^0_n$, $\omega^\mathrm{g}_{\bullet\, n}$, and $\omega^\mathrm{th}_{\bullet\, n}$ grows with particle size, especially for the dipolar mode, reflecting spectral distortions due to absorption and radiative broadening. For example, the dipolar mode in the $50$~nm particle exhibits a $>200$~meV shift between its passive and emission frequencies.

Second, the threshold gain increases dramatically with size for the dipolar mode, while remaining more stable for the quadrupolar mode and even decreasing slightly for the octupolar one. These results underline how larger particles favor higher-order resonances from an energetic standpoint.

These behaviors are further clarified in the \emph{Extinction Cross Section} section, where the modal interplay under gain is examined. Within the parameter space explored here, we find that dipolar and quadrupolar modes can be independently driven to emission with minimal interference—consistent with and extending the findings of Ref.~\cite{Recalde:2023}. However, selectively exciting the octupolar mode is more challenging. For small nanoparticles (10–30~nm), the octupolar and quadrupolar thresholds overlap, and the quadrupolar mode tends to dominate as gain increases, leading to competition and loss of selectivity. Only at larger sizes (e.g., 50~nm), where modes are more spectrally separated, can octupolar emission occur without amplifying lower-order modes.

This characterization demonstrates that while mode-specific emission is possible, it requires careful design of both the gain profile and the nanoparticle size to mitigate modal competition and achieve controlled plasmonic lasing.

\section*{Conclusions}

In this work, we systematically investigated the emission behavior of metal nanoparticles embedded in an infinite gain medium by applying Mie Scattering Theory beyond the quasi-static regime. Focusing on the first three multipolar modes ($n = 1, 2, 3$), we introduced and implemented a robust numerical framework to determine, for each mode and particle size, the optimal gain center frequency $\omega^\mathrm{g}_n$, the minimal gain threshold $G^\mathrm{th}_n$, and the corresponding emission frequency $\omega^\mathrm{th}_n$. These quantities were obtained by identifying the complex singularities of the electric Mie scattering coefficients $a_n(\omega)$, under finely tuned gain conditions.

Our analysis revealed a strong dependence of both emission frequency and threshold gain on the nanoparticle radius and mode order. In particular, we showed that for larger particles, higher-order modes tend to require lower gain to reach the emission regime, owing to their narrower linewidths and weaker radiative losses. This finding aligns with previous reports based on time-domain simulations, but here it is derived from a fully analytical steady-state treatment.

By computing extinction spectra across a range of gain configurations and particle sizes, we also demonstrated how the collective optical response of the nanoparticle reflects the interplay between different modes. We observed that while emission can be selectively induced in a target mode by centering the gain at $\omega^\mathrm{th}_n$, spectral overlap can lead to unintended amplification of neighboring modes—particularly when the gain linewidth is comparable to or broader than the modal separation.

Notably, we identified a particle size regime (e.g., $r = 50$~nm) where the spectral separation is sufficient to allow clean, mode-selective excitation of the octupolar resonance without interference from lower-order modes. This opens promising perspectives for the design of plasmonic nanolasers and spasers with enhanced modal control.

Future work may extend this approach to more complex geometries (e.g., core-shells or non-spherical particles), as well as to time-resolved or nonlinear models capable of capturing the full emission dynamics beyond the steady-state threshold.

\section*{Acknowledgments} 

M.~A.~I., O.~M.~M., M.~I., and A.~V. acknowledge funding by the European Union (NextGeneration EU), through the MUR–PNRR project PE0000023–NQSTI, the PRIN2022 project “Cosmic Dust II” (grant No. 2022S5A2N7), and research carried out within the project “Space It Up” funded by ASI and MUR – Contract No. 2024-5-E.0 – CUP No. I53D24000060005.

The manuscript was written through contributions of all authors. All authors have given approval to the final version of the manuscript.

\section*{Conflict of Interest}
The authors declare no conflicts of interest.

\section*{Data Availability}
The data that support the findings of this study are available from the corresponding author upon reasonable request. Numerical codes used in this study are available from the authors upon request.

\section*{Ethics Statement}
This work does not involve human participants, animal experiments, or sensitive data.

\bibliographystyle{iopart-num}
\bibliography{updated_2025}
\end{document}